\documentclass[12pt,preprint]{aastex}
\usepackage{epsfig,lscape} 
\usepackage[dvipdfm]{hyperref} 
\usepackage{natbib}
\usepackage[caption=false]{subfig}
\usepackage{lscape}
\newcommand{\dfn}{\hbox{$D_{\rm n}$(4000)}}
\newcommand{\ha}{\hbox{H$\alpha$}}
\newcommand{\hb}{\hbox{H$\beta$}}
\newcommand{\oii}{\hbox{[O\,{\sc ii}]}}
\newcommand{\oiii}{\hbox{[O\,{\sc iii}]}}
\newcommand{\hg}{\hbox{H$\gamma$}}
\newcommand{\nii}{\hbox{[N\,{\sc ii}]}}
\begin{document}

\title{The metallicity evolution of blue compact dwarf galaxies from {the} intermediate redshift to {the} local Universe}

\author{Jianhui Lian\altaffilmark{1}, Ning Hu\altaffilmark{1}, Guanwen Fang\altaffilmark{2}, Chengyun Ye\altaffilmark{1}, Xu Kong\altaffilmark{1}}
\email{ljhhw@mail.ustc.edu.cn; xkong@ustc.edu.cn}
\altaffiltext{1}{CAS Key Laboratory for Research in Galaxies and Cosmology, Department of Astronomy, University of Science and Technology
of China, Hefei, Anhui 230026, China} 
%\altaffiltext{2}{CAS Key Laboratory for Research in Galaxies and Cosmology, Hefei, Anhui 230026, China}
\altaffiltext{2}{Institute for Astronomy and History of Science and Technology, Dali 
University, Dali 671003, China}
%Department of Astronomy, University of Science and Technology of China, Hefei 230026, China\\
%Key Laboratory for Research in Galaxies and Cosmology, Chinese Academy of Sciences, Hefei 230026, China}

%\email{ljhhw@mail.ustc.edu.cn (JHL); xkong@ustc.edu.cn(XK)}

\begin{abstract}
We present {oxygen} abundance measurements for {74} blue compact dwarf (BCD) galaxies {in the} redshift {range} {in $[0.2,0.5]$}
{using the strong-line method}.
The spectra of these objects are taken {using} Hectospec on {the} Multiple Mirror Telescope (MMT).
{More than} half of these BCDs {had} 
dust attenuation corrected {using the} Balmer decrement method.
For comparison, we also {selected} a sample of {2023} {local} BCDs from {the} Sloan Digital Sky Survey (SDSS) database.
Based on the local and {intermediate-$z$} BCD {samples}, we {investigated} the {cosmic} evolution of 
{the} {metallicity, {star-formation} rate (SFR), and \dfn\ index.}
%R23 and O32
%atio which are both indicators of metallicity and ionization parameters.}
Compared {with} local BCDs,
{the intermediate-$z$} BCDs {had a} systematically higher R23 ratio {but} similar 
O32 ratio. 
Interestingly, no significant {deviation in} {the} {mass-metallicity {(MZ)} relation} {was} found {between the 
intermediate-$z$ and local BCDs}.
{Besides the metallicity, the intermediate-$z$ BCDs also {exhibited an} SFR distribution {that was consistent} with 
local BCDs, suggesting {a} weak dependence on redshift.}
{The} intermediate-$z$ BCDs {{seemed} to be younger than {the} local BCDs with lower 
\dfn\ index {values}.}
{The insignificant deviation in {the} mass-metallicity and mass-SFR {relations} between intermediate-$z$ and local BCDs {indicates} that 
{the} relations between {the} global parameters of {low-mass} compact galaxies may be universal.}
These results {from} low mass {compact} galaxies {could} {be used to place} important {observational} constraints 
{on} galaxy formation and evolution models.
\end{abstract}


\section{introduction}
Metallicity is one of the most important {parameters} of galaxies {and provides} information 
{on their} past evolution.
Many physical {processes}, such as star formation {and} gas inflow and outflow,
involve {the assembling {of mass and}} the metal enrichment {of a galaxy}.
A strong correlation between total mass (gas and stellar mass) and metallicity was found in the 1970s \citep{lequeux1979}.
Later on, \citet{tremonti2004} confirmed this relation based on a sample of $\sim$ 50,000 star forming galaxies at $z \sim$ 0.1
from {the} Sloan Digital Sky Survey (SDSS; \citealt{abazajian2004}). 
Galaxies with lower {stellar} mass {tend to be less metal-enriched}.
They also found that the dispersion of {the} mass-metallicity (MZ) relation is {lower} than that of luminosity-metallicity
relation and concluded that the former is more {fundamental}. 
With deep spectroscopic {surveying of} distant galaxies, it is {possible} to investigate the MZ relation in the early universe.
Recently, the MZ relation at intermediate and high redshift {has} been extensively explored 
\citep{savaglio2005,erb2006,maiolino2008,zahid2011,yuan2013}. 
When the MZ {relations} at different redshifts {are compared}, galaxies at higher redshift tend to be more metal-poor
at a fixed {stellar} mass. This trend was often regarded as {the} cosmic evolution of {the} MZ relation \citep{maiolino2008}.
{The evolution of {the} MZ relation was also found when {comparing the analogs} of high redshift galaxies with local {star-forming} galaxies \citep{hoopes2007,ljh2015}.}
However, such {a} comparison may be significantly affected by the different sample selection criteria and different methods for 
metallicity and stellar mass determination \citep{Izotov2015}.

The MZ relation is an important
probe {for} the balance between gaseous inflows, outflows, and star formation. One possible explanation {for} the
MZ relation is supernova-driven winds, which could remove the metal-enriched gas and
are more effective in {low-mass} galaxies with {a} shallow gravitational potential. It is also suggested by many {studies} 
that outflows are prevalent in {low-mass} galaxies \citep{Izotov2006a,Martin2012}. However,
the properties of these gaseous flows are poorly constrained and the physical driver of {the} MZ relation is still
under debate. A better understanding of {the} MZ relation at low mass are important to fully understand the MZ
relation and how galaxies {evolve at the} early stage. There are many {studies of} the local MZ relation (or
luminosity-metallicity relation) of {low-mass} galaxies \citep{Lee2006,Zhao2010,Berg2012}.
However, outside the local universe, the MZ relation at stellar {masses} below $10^9 M_{\odot}$ 
{is not {yet} well established}. 

Blue compact dwarf (BCD) galaxies are faint ($M_{\rm B}>$-17 mag) objects and have distinctive compact morphology. 
Strong emission lines in optical spectra indicate that they are undergoing
active star formation \citep{sergent1970,kunth2000,kong2002}. 
According to hierarchical {galaxy-formation scenarios}, dwarf galaxies are part
of the building blocks for massive galaxies in the early universe. The gas consumption timescale of BCDs, given the
current SFR, is much shorter than
the age of the universe \citep{kong2004}. In this work, we consistently select BCDs at {the} {intermediate-$z$ 
and local universe} and
{investigated} the possible cosmic 
evolution of {the} MZ relation at {the low-mass} end. 

This paper is organized as follows.
In Section 2, we describe the selection of BCDs at different redshifts and the spectroscopic observation
{of} BCDs at intermediate redshift. The measurements {for} emission {lines} and physical properties
such as stellar {mass}, SFR, {\dfn\ }index and metallicities are described in Section 3.
{We then} provide the MZ relation of BCDs at different redshifts {using}
the same metallicity determination method in
Section 4. In this part, we also investigate the {possible evolution of {the} SFR and \dfn\ index.}
%dependence of MZ relation on SFR and its possible cosmic evolution.
Finally, we present a summary in Section 5.
Throughout this paper, we adopt the cosmological parameters $H_0=70\, {\rm km s^{-1} Mpc}^{-1}$, $\Omega_{\Lambda}=0.73$ 
and $\Omega_{\rm m}=0.27$.

\section{Sample selection and observation}
\subsection{BCDs at $z \sim 0.03$}
There are many definitions of BCD in the literature \citep{thuan1981,kong2002,gilde2003}. 
Generally, BCDs are characterized by their blue color, compact morphology, and low luminosity or stellar mass. \citet{gilde2003} proposed  
quantified classification criteria to select BCDs {as follows}: 1) blue color with $B-R < 1$; 2) compact 
with peak surface brightness in the $B$ band $\mu_{\rm B}<22~$mag ${\rm arcsec}^{-2}$; {and} 3) faint with $M_{\rm K}<-21$~mag. Later on, \citet{sanchez2008}
used a set of similar criteria based on SDSS filters (see Table 3 in \citealt{sanchez2008}). 
According to the definition of BCD in \citet{sanchez2008},
we {selected} galaxies from SDSS DR12 \citep{alam2015} that {satisfied} the following criteria as our BCD sample in the local {universe}: 
1) blue color with $\mu_{\rm g}-\mu_{\rm r}<0.43$~mag~${\rm arcsec^{-2}}$, where $\mu_{\rm g}$ is the mean surface brightness within $r_{50}$ in the $g$ band; 
2) compact with $\mu_{\rm g}<21.83-0.47(\mu_{\rm g}-\mu_{\rm r})$; 3) dwarf with stellar mass $M_*<10^9\ {\rm M_{\odot}}$; 
4) stellar mass error $\triangle{\rm m}<0.25$ dex; 5) redshift $z<0.05$; 6) signal-to-noise ratio (SNR) of \oii$\lambda$3727, 
\hb, \oiii$\lambda\lambda$4959,5007 emission line
above 5. All {of} the measurements {were} taken from the MPA-JHU catalog 
\footnote{http://skyserver.sdss.org/dr12/en/help/browser/browser.aspx} \citep{kauffmann2003,brinchmann2004}.
Since {the error of} {the} stellar mass is {not directly given} in the MPA-JHU catalog, we {estimated} the error by 
calculating half of the difference between the 84th percentile 
and {the} 16th percentile {probability distribution of} {the} stellar mass (i.e. {the} lgm\_{tot}\_{p84} and lgm\_tot\_p16 index) in the MPA-JHU catalog. 
Contamination of {the} HII regions {in nearby galaxies} {was} excluded by visually inspecting the SDSS image.
Finally, {2023} galaxies {were} selected as our local BCD sample with a median redshift of $0.03$\
and a median {equivalent width (EW) \hb} of 14.9 \AA. {Among them, 262 BCDs {had} \oiii$\lambda$4363 line detected at {the} 5$\sigma$ level.}

\subsection{BCDs in the COSMOS field}
To study {{low-mass} galaxies} outside the local {universe}, we {selected} a BCD sample at intermediate redshift in the 
Cosmic Evolution Survey (COSMOS) {deep} field and then {performed} follow-up 
spectroscopic observation using Hectospec on {the Multiple Mirror Telescope (MMT)}. 
We {selected} the BCD sample from a public $K_{\rm s}$ selected catalog of {the} COSMOS field \citep{muzzin2013}.
COSMOS is a Hubble Space Telescope (HST) 
Treasury Project to survey a {two-square-degree} equatorial field with comprehensive observations
from radio to X-ray.
We {searched} for BCDs in a {one-degree} diameter circle centered at ${\rm RA=150.1152, Dec=2.3420\ (J2000)}$. 
Only galaxies at redshift lower than 0.7 {were} selected because the \oiii$\lambda$5007 line at higher redshift
will be redshifted beyond  
the wavelength coverage of {the} Hectospec/MMT.
For the dwarf criteria, we {used the} {stellar mass $M_{*}$}, which could be accurately measured based on
27 broad- and intermediate-band {filters}.%, rather than luminosity in the $K_{\rm s}$ band. 
To reliably estimate the peak surface brightness,{ high-resolution imaging} is needed. Since only one HST band at 0.814 $\mu$m is available, we
{obtained} an image cut for each dwarf galaxy and then
{calculated} the peak surface brightness in the $F814W$ band using SExtractor \citep{bertin1996}. 
A correction factor of $(1+z)^4$ {was} applied to the surface brightness to account for the redshift dimming effect.
To roughly cover the restframe $U-B$ color of {the} intermediate-redshift objects, we use the observed $u-i$ color based on {the} SDSS filter system.
The distribution of the $u-i$ color of dwarf galaxies in the catalog has a Gaussian 
profile with a median {of} 1.3, rather than {the} bimodal distribution found for 
massive galaxies \citep{baldry2004}. We {decided} to select the bluer half dwarf galaxies to exclude possible contamination {by} red early-type 
dwarf galaxies. 
According to the Mass-Excitation diagram \citep{juneau2011},
the AGN fraction at $M_*<10^9 M_{\odot}$ is negligible.
Therefore,
we {did} not exclude AGN contamination from the BCD sample.
The final criteria for BCDs in the COSMOS field {were}: 
1) $u-i<1.3$; 2) $\mu_{\rm F814W, peak}<22$~mag ${\rm arcsec}^{-2}$; 3) $M_{*}<10^9M_{\odot}$; 4) $z<0.7$.
%; 5) located in the $1\,^{\circ}$ diameter circle.   
In total, $\sim$ 1400 objects {were} {selected} as {a parent sample of} BCD. {We then ran} the {\sc xfitfibs} program 
which {was} developed for Hectospec 
to allocate the fibers to the objects. 
Every {pair of} objects should be separated by at least 22$''$ to avoid fiber collision.
In the end, 180 BCDs with a median redshift of 0.39 and 28 
sky background regions {were} assigned {to} a fiber.

The {observations} of these BCDs {using the} {Hectospec/MMT} {were} carried out in 2015 February.
Hectospec is a moderate-resolution, multi-object spectrograph, fed {by} 300 fibers. 
Each fiber has a diameter of 1.5$''$ (corresponding to 5 kpc at $z=0.2$) and covers most of the light
of BCDs at intermediate redshift.
The field of view is a circle {with {a} diameter of $1^{\circ}$}. We {chose} the 270-line grating with {a} resolution 
of 1.2 \AA\ per pixel to achieve wider wavelength coverage from 3650 \AA\ to 9200 \AA\
{and {a} wavelength resolution of $\sim$ 6 \AA}. 
Figure 1 shows the restframe spectra smoothed {using} a boxcar of {five} pixels for the first four COSMOS BCDs as listed in Table 1.
The fiber spectra {were} reduced using the 
HSRED reduction pipeline. The data {were} flux calibrated using the observation of one standard star, PG-1545+035, taken {1} week before the science 
observation. Sky subtraction {was} achieved by averaging the spectra of sky regions from the same exposure.
With an exposure time of 4 hours, 141 objects {had} \oiii$\lambda\lambda4959,5007$ double lines detected at a level above $3\sigma$. Figure 2 shows the 
comparison {between the} spectroscopic redshift from this observation {and} {the} photometric redshift in the catalog of \citet{muzzin2013}.
There are eight objects that have spectroscopic redshift {from {the} zCOSMOS survey} \citep{lilly2009} and {they} are marked as {filled} circles in Figure 2.
It can be seen that the photometric redshifts are roughly consistent with {the} spectroscopic {redshifts} with an offset of 
-0.007$\pm$0.112 dex.% and dispersion of 0.112 dex {in average}.

\begin{figure*}
 \includegraphics[width=\textwidth]{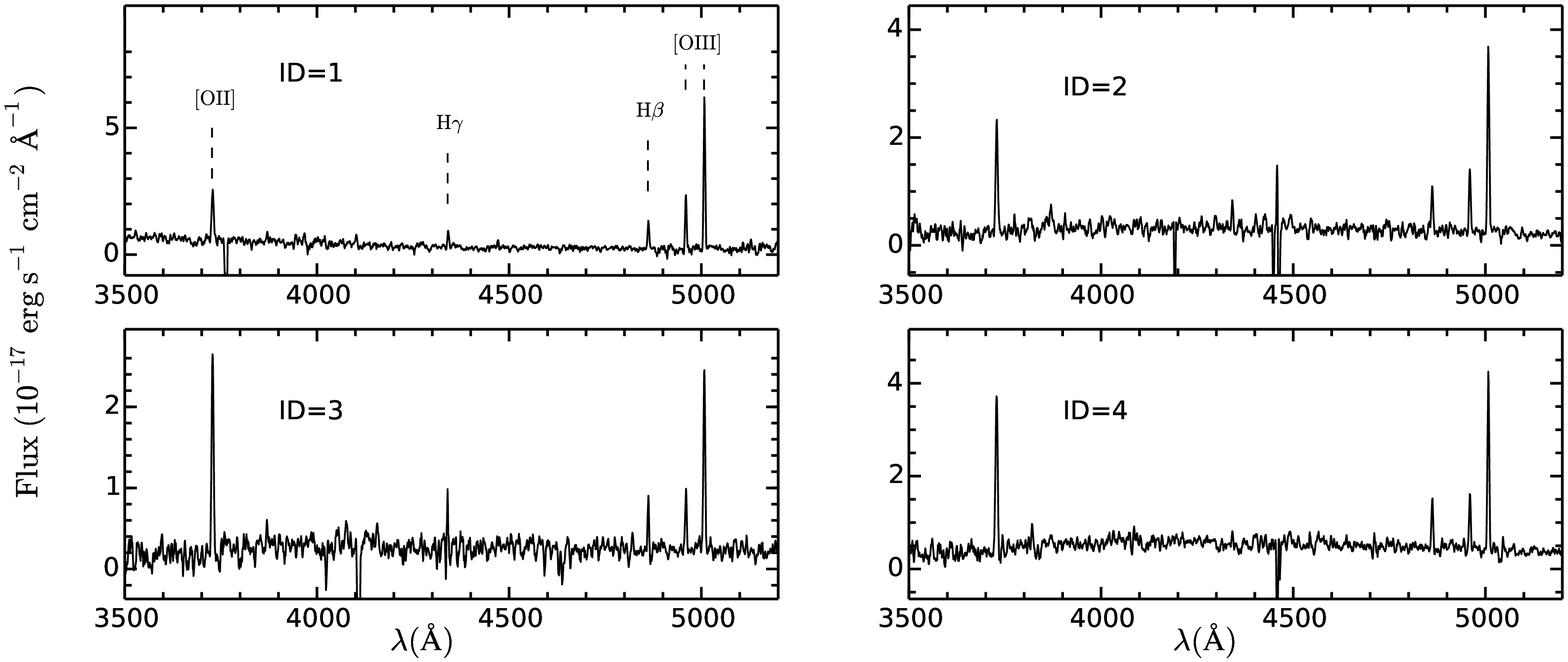}
 \caption{Hectospec spectra {smoothed {using} a boxcar of {five} pixels for} the first four COSMOS BCDs as listed in Table 1.
Each panel is marked with the corresponding object ID.}
\label{figure1}
\end{figure*}

\begin{figure*}
 \centering
 \includegraphics[width=\textwidth]{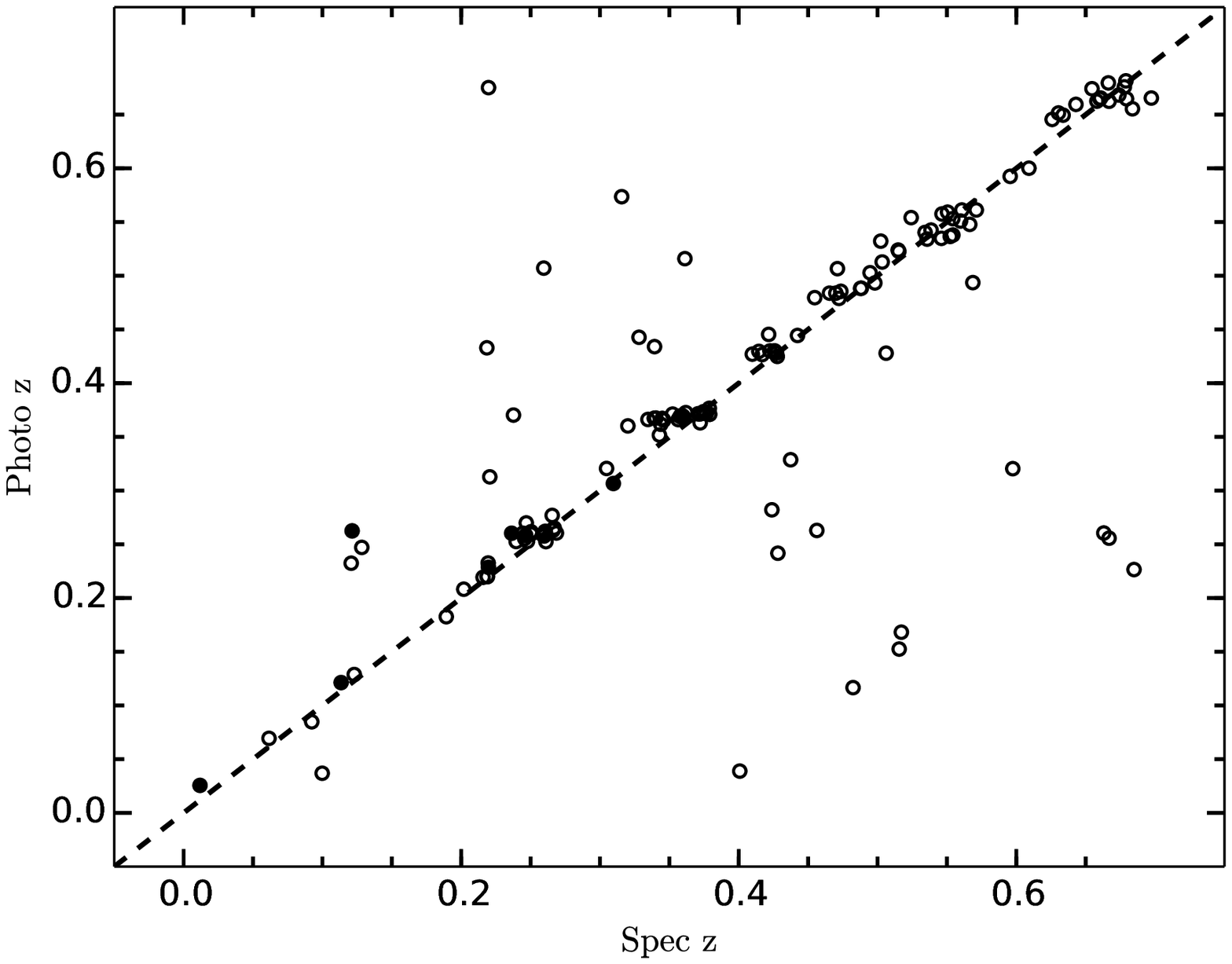}
 \caption{Comparison between {spectroscopic} and photometric redshift for COSMOS BCDs. {Filled} circles are BCDs with spectroscopic
redshift from \citet{lilly2009}. Dashed line indicates equality between the x- and y-axis values.}
\label{figure2}
\end{figure*}

\section{Emission-line Measurements}
Emission-line measurements of the SDSS BCD sample {were} taken from the MPA-JHU catalog. 
{For BCDs in the COSMOS field, after subtracting the spectra continuum, 
we {obtained} line fluxes by fitting Gaussian profiles to each line using {the} {\sc MPFIT} routine \citep{markwardt2009}}
{Since most} of {the} {COSMOS} BCDs are faint, their spectra {continua} are marginally detected. 
To {subtract {the} spectra continuum and estimate the} significance of emission lines, we {followed} the method in \citet{ly2014}.
Firstly, {for each line}, we {picked} a 200 \AA-wide {spectrum}, excluding regions affected by sky lines and nebular emissions. 
The spectra continuum is assumed to be the median of this {spectrum}.
%After subtracting the continuum, we fit Gaussian profiles to each line using MPFIT routine \citep{markwardt2009}. 
{We then computed the} SNR by dividing the {line} flux by:
\begin{equation}
 {\rm Noise} = \sigma \times l \times \sqrt{N_{\rm pixel}},
\end{equation}
, where $\sigma$ is the standard deviation of the 200 \AA-wide spectra, $l$ is the spectral resolution and 
$N_{\rm pixel} = 5\sigma_{\rm G}/l$ (also see {equation} 3 in \citet{ly2014}).
$\sigma_{\rm G}$ is the Gaussian width of the emission line. 

To derive {the} metallicity {using the} R23 method (R23 = log((\oii$\lambda$3727+\oiii$\lambda \lambda$4959,5007)/\hb)), we {selected} a 
subsample of {74} COSMOS BCDs {whose strongest emission lines had an SNR} above 3$\sigma$ 
{and {with a redshift in the range} [0.2,0.5]} as {in} our COSMOS BCD sample.
%{These COSMOS BCDs have a median redshift of 0.34$\pm$0.08.}
{Since the diameter of {the} MMT is 2.6 times larger than that of {the} SDSS and 
the exposure time is 5.3 times longer, the SNR criteria of {the} SDSS BCDs {were} actually higher than {for} the COSMOS BCDs by a 
factor of 10, which is roughly proportional to 1/$z$. 
Therefore, these two BCD samples should be comparable with consistent {emission-line} selection criteria.}
All of the {line} fluxes of the COSMOS and SDSS BCDs {were} corrected for {galactic} foreground extinction. 
Since the spectra {continua} are marginally detected, we {were} not able to accurately determine the Balmer absorption.
\citet{zahid2011} {estimated} the Balmer absorption, which is $\sim$ 1 \AA\,  
in {EW}, for {star-forming} galaxies at $z\sim0.8$. 
According to their result, we {applied} a correction of 1 \AA\, in {EW} {for {the} Balmer absorption} to our Balmer {emission-line} measurements.
It should be noted that {such} correction {did} not significantly affect our 
results, {as it} is small compared to the EW(\hb) of COSMOS {BCDs, with a median value of} 30.2 \AA. 

Since {the} R23 indicator is sensitive to dust attenuation, it is important to {reliably} determine the 
intrinsic extinction.
In this work, we {used the} Balmer decrement method to determine the 
dust attenuation assuming a Calzetti extinction law \citep{calzetti2000}.
Since the spectra are affected by contamination of second-order light at {wavelengths} longer than $\sim 8200$\AA,  
the \ha\ line can not be detected or reliably flux calibrated for BCDs at $z>0.25$.
For consistency, we {used} the \hb/\hg\ ratio to estimate the dust attenuation assuming {an} intrinsic ratio of 2.14. 
However, {of the COSMOS BCD {samples}}, only {38} BCDs {had an} SNR of {the} \hg\ line above $3\sigma$. 
For the rest of {the} {36} {COSMOS} BCDs, we {used} the color excess $E(B-V)$ from SED fitting to roughly
estimate the intrinsic extinction.
It should be noted that the nebular extinction {was} higher than {the} stellar extinction.
Although the exact relation between {the} nebular and the stellar extinction is under debate,
we {adopted} a factor of 2.3 to transfer the stellar extinction from {the} SED {fit} to {the} dust extinction of {the} emission lines
\citep{calzetti2000,cresci2012}. 
In the rest of this paper, we will {distinguish between the} COSMOS BCDs {using} these two {extinction} determination {methods}.
The extinction-corrected {emission-line} fluxes are listed in Table 1.
We do not provide an {estimate} of {the} uncertainties of {the} extinction from {the} SED {fit} which {was}
strongly {dependent upon the} other SED fitting parameters.
{Of the COSMOS BCDs with dust extinction corrected {using the} Balmer decrement method, {two-thirds exhibited} 
negligible dust extinction ($A_{\rm V}=0$).}
%Since about two thirds of the Balmer decrement corrected BCDs have negligible
%dust extinction ($A_{\rm V}=0$), 
{In these cases,} the uncertainties of {the} dust extinction can not be precisely determined.
Therefore, we {did} not propagate the uncertainties {of {the}
dust extinction} to {that of} the extinction-corrected {emission-line} fluxes.

{The extinction E(B-V) would be negative for galaxies with {an} \hb/\hg\ ratio {of} less than 2.14. 
In these cases, we manually set the extinction to 0 {since the dust extinction
{seemed} to be negligible, which {was} indicated by the {relatively} low \hb/\hg\ ratio}.
{Figure 3} shows the R23 and O32 (O32=log(\oiii$\lambda$5007/\oii$\lambda$3727)) {ratios} versus {the} stellar mass.
{The grey} dots represent {the} SDSS BCDs and {the} red circles are {the} COSMOS BCDs.
{The filled} circles are {the} BCDs with dust attenuation corrected {using the}
Balmer decrement method. 
{The R23 ratio of {the} COSMOS BCDs seems to be systematically higher than that of {the} SDSS BCDs at $M_*>10^8 M_{\odot}$,
while the O32 {ratios} of the two BCD samples are rather similar.}
\begin{figure*}
 \centering
 \includegraphics[width=\textwidth]{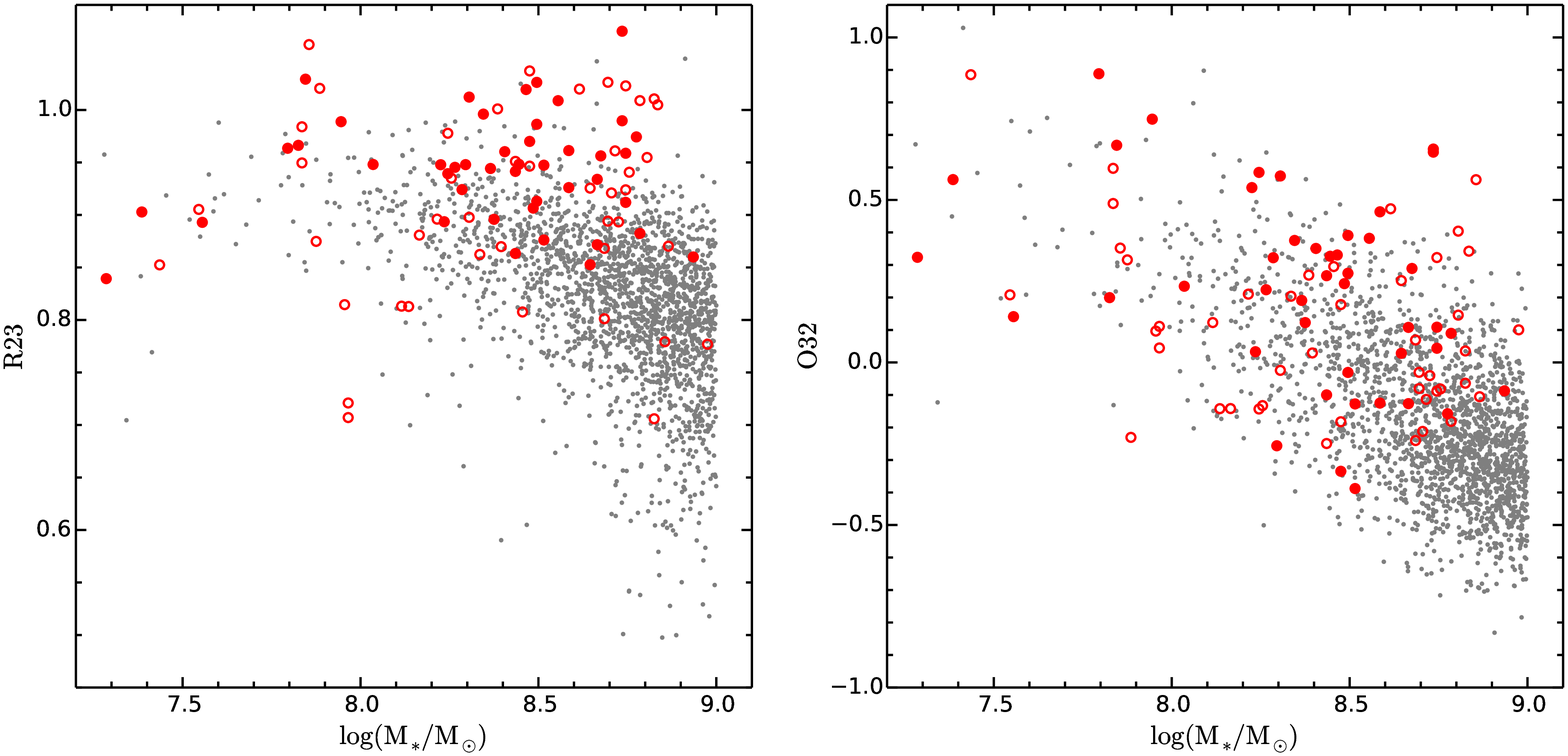}
 \caption{R23 ratio (left panel) and O32 ratio (right panel) distribution versus {the} stellar mass. 
 Grey dots are {the} SDSS BCDs and red circles are {the} COSMOS BCDs. 
 The {filled} circles represent COSMOS BCDs with dust attenuation corrected {using the} Balmer decrement method.}
\label{figure3}
\end{figure*}

\section{Physical properties Determination}

The {total} stellar mass of {the} SDSS BCD sample {was} taken from the MPA-JHU catalog where a Kroupa IMF \citep{kroupa2001} was assumed. 
{The} MPA-JHU team obtained the stellar mass based on \citet{bc03} (hereafter BC03) population synthesis model and 
{assuming} {an} {exponentially-declining}
star formation history.
The stellar mass of {the} COSMOS BCD sample {was} determined by fitting {the} stellar population templates to 27
emission-line-corrected broad- 
and intermediate-band {filters using} the Fitting and Assessment of Synthetic Templates (FAST; \citealt{kriek2009}) code.
The original {photometric were} taken from the {photometry} catalog of \citet{muzzin2013}. 
Since BCDs {typically} have high emission-line {EW}, we {needed} to correct the 
broad- and intermediate- band photometry for the contribution from emission lines.
For each galaxy, we {generated} a redshifted spectrum with emission-lines and zero {continua}. 
{We then convolved} these spectra with the 27 filter bandpasses to determine the flux {excesses}, and {then} {removed} the excesses from {the}
original photometry.
For the objects without reliable \ha\ detection, we {estimated} the contribution {from} \ha\ 
based on \hb\ {measurements}, assuming {an $\ha/\hb$ ratio of 2.86.}
To {allow a direct comparison {with} local BCDs}, we {chose} the BC03 population synthesis model and {adopted} {an}
{exponentially-declining star-formation} history to perform the SED {fit}. 
However, there is no Kroupa {IMF-based} BC03 model available in the FAST default {settings}.
Therefore, we {adopted} a Chabrier IMF \citep{chabrier2003} and then {applied} a correction factor of 1.06 to transfer to {the}
Kroupa IMF based stellar mass.

{The} SFR {of {the} BCDs {was} determined} using the relation between {the} {extinction-corrected} \ha\ luminosity 
and {the} SFR in \citet{hao2011}. Since most COSMOS BCDs have no \ha\ detection,
\hb\ luminosity {was} used instead, assuming {the} intrinsic ratio of \ha/\hb\ to be 2.86.
{In the burst scenario, \ha\ and \hb\ {EWs} strongly decline with the starburst age
\citep{leighterer1999, Izotov2015}. Therefore, to perform {a} fair comparison, we {corrected the} 
SFR to the initial stage of burst according to {equation (5)} in \citet{Izotov2015}.
However, it should be noted that the correction will be overestimated when there are old {populations} 
present in the galaxy. The overestimation will be stronger for galaxies with lower EW(\hb).}
{The SFR {measurements} of SDSS BCDs {were} corrected for {the} aperture effect, which {was} estimated {using} the difference 
between the model and {the} fiber magnitude at {the} $u$ band as $10^{(u_{\rm fiber}-u_{\rm total})/2.5}$. The 
aperture corrections are small, with a median value of 2.8. 
We {did} not apply aperture corrections for {the} COSMOS BCDs because the fiber {coverage} could be higher 
than the local BCDs by an order of magnitude and the corrections must be negligible.}
%{For SDSS BCDs, we apply aperture corrections of }
We {calculated} the 4000 \AA\ break [i.e. \dfn\ index] based on the definition of \citet{balogh1999}.
%which is also used by \citet{kauffmann2003}. 
Since the spectra continuum are marginally detected, {42 of the COSMOS BCDs have} reliable measurements 
of \dfn\ index with {3$\sigma$ uncertainty
less than 0.11}.

There are many methods {for obtaining} the metallicity of an emission-line galaxy. 
One of the most direct is the so-called `$T_{\rm e}$-method', which {utilizes} the anticorrelation between {the} metallicity and 
the electron temperature $T_{\rm e}$. The electron temperature can be derived from the ratio of the auroral to {the}
nebular emission lines. However, the auroral lines, such as \oiii$\lambda$4363, are always very weak. 
Therefore, many empirical and theoretical calibrations are proposed. 
Although these calibrations are all based on {strong-line} ratios, the difference could be as large as 0.7 dex
\citep{shi2005,kewley2008}. The origin of this difference is still not fully understood. 

{In this work, we {obtained} the metallicity {using} the relation between {the} metallicity and {the} R23 ratio
and {the O32 ratio}.}
Since there are two metallicity solutions for one R23 value, extra information {is} needed to 
discriminate the upper or lower metallicity solution. The diagnostics for distinction should be sensitive to 
metallicity.
The most widely-used {is} the ratio of \nii$\lambda$6584/\ha\
or \nii$\lambda$6584/\oii$\lambda$3727 \citep{kewley2008}. However, the \nii$\lambda$6584 is redshifted to 
{the} infrared at intermediate redshift and {sky lines} could strongly contaminate this weak emission line. 
Therefore,
other methods without the information of \nii$\lambda$6584 are needed to break the R23 degeneracy. 
{Other parameters, such as {the} O32 
line ratio and stellar mass, are proposed to break the R23 degeneracy 
\citep{maiolino2008,henry2013}. 
\citet{maiolino2008} (hereafter M08) {first} calibrated the R23- and O32-metallicity {relations} by combining the ${\rm T_{e}}$ method and 
photoionization models.
Then, metallicities in their work {were} derived by minimizing the difference from 
the {expectations of the} R23 and O32 measurements simultaneously.
\citet{henry2013} determined metallicities using the R23 calibration from \citet{kobulnicky2004} 
but utilized stellar mass to break the R23 degeneracy.
For a subsample of 262 local BCDs which have \oiii$\lambda$4363 {detected}, we {compared the} 
metallicities {determined} {using} these two methods {to that} {derived from} the direct ${\rm T_{e}}$ method {\citep{Izotov2006b}}.
The M08 method {shows a stronger correlation with the Te method than the method used in Henry et al. (2013).}
We also derived {the} metallicities for local BCDs {using} N2 method \citep{pp04}, ${\rm 12+log(O/H) = 8.9+0.57*log(\nii/\ha)}$, which is monotonic and insensitive to dust attenuation.
{Metallicities derived {using} the} M08 method also show a stronger correlation with {that derived {using} the} N2 method. 
In total, we consider the M08 method to be more robust than the KK04 calibration with stellar mass for breaking {the} R23 degeneracy.
Therefore, in this work, we {adopted} the method from \citet{maiolino2008} to derive the metallicity of BCDs.}

{We {repeated} the metallicity determination process 1000 times
while the {emission-line} fluxes {varied} based on their uncertainties.
The median value of the 1000 measurements {was} the final metallicity and the 
median of {the} absolute deviation from the final metallicity {was} the error of metallicity.}
The obtained measurements {of all the parameters mentioned above} are listed in Table 2.

\section{Results}
\subsection{Mass-metallicity relation}
The MZ {relations} at intermediate and high redshift have been extensively explored. 
However, most of the work {has} focused on massive {star-forming} galaxies, which can be reached at {high} redshift. 
{Recently, \citet{henry2013} {were the first to derive} the low-mass, {intermediate-$z$} MZ relation {based on}
26 emission-line galaxies at $z \sim$ 0.6 - 0.7. 
Most of the galaxies in their sample {had a} stellar mass higher than $\sim 10^{8.4} {\rm M_{\odot}}$.
They obtained metallicities using the R23 calibration from \citet{kobulnicky2004}.
Based on the MZ relation, they utilized stellar 
mass to {chose} the appropriate branch of {the} R23 solution. % based on MZ relation.
They found {that} the metallicity is {typically}
lower than that at $z\sim 0.1$ by about 0.12 dex at a fixed mass. 
However, {the dust extinction in their work was estimated} using SED fitting, which may suffer 
large uncertainties.}

In this work, we {aimed} to study the {intermediate-$z$} MZ relation at {the} {lower stellar mass} end {with {a} much larger 
and {more} complete sample}.% down to $\sim 10^{7.3} {\rm M_{\odot}}$
%by taking spectra of 180 BCDs with Hectospec on MMT. 
{Figure 4} shows the MZ relation of {the} {intermediate-$z$ COSMOS and local SDSS} BCDs .% with reliable emission line detection which allows metallicity determination.
The {gray} dots {represent} the SDSS BCDs.% in the mass-metallicity diagram. 
The red {filled} circles are {the} COSMOS BCDs with dust {extinction} corrected {using the}
Balmer decrement method, while the empty circles {represent the} correction {from} SED fitting.
{The dashed line is the best-fitting MZ relation {to} {the} local BCDs as:
\begin{equation}
12+{\rm log(O/H)}=4.55(\pm0.10)+0.46(\pm0.01)\times {\rm M_*} .
\end{equation}
}
%We divide our COSMOS BCD sample into two redshift bins with $z$ in [0.2,0.4] and [0.4,0.7] 
%and separately plot in the two panels of {Figure 5}.
All stellar mass {values were} calibrated to be consistent
with {the} Kroupa IMF and {the} metallicities with calibration from \citet{maiolino2008}.
%KK04 calibration using the equations in
%\citet{kewley2008}. 

\begin{figure*}
 \centering
 \includegraphics[width=\textwidth]{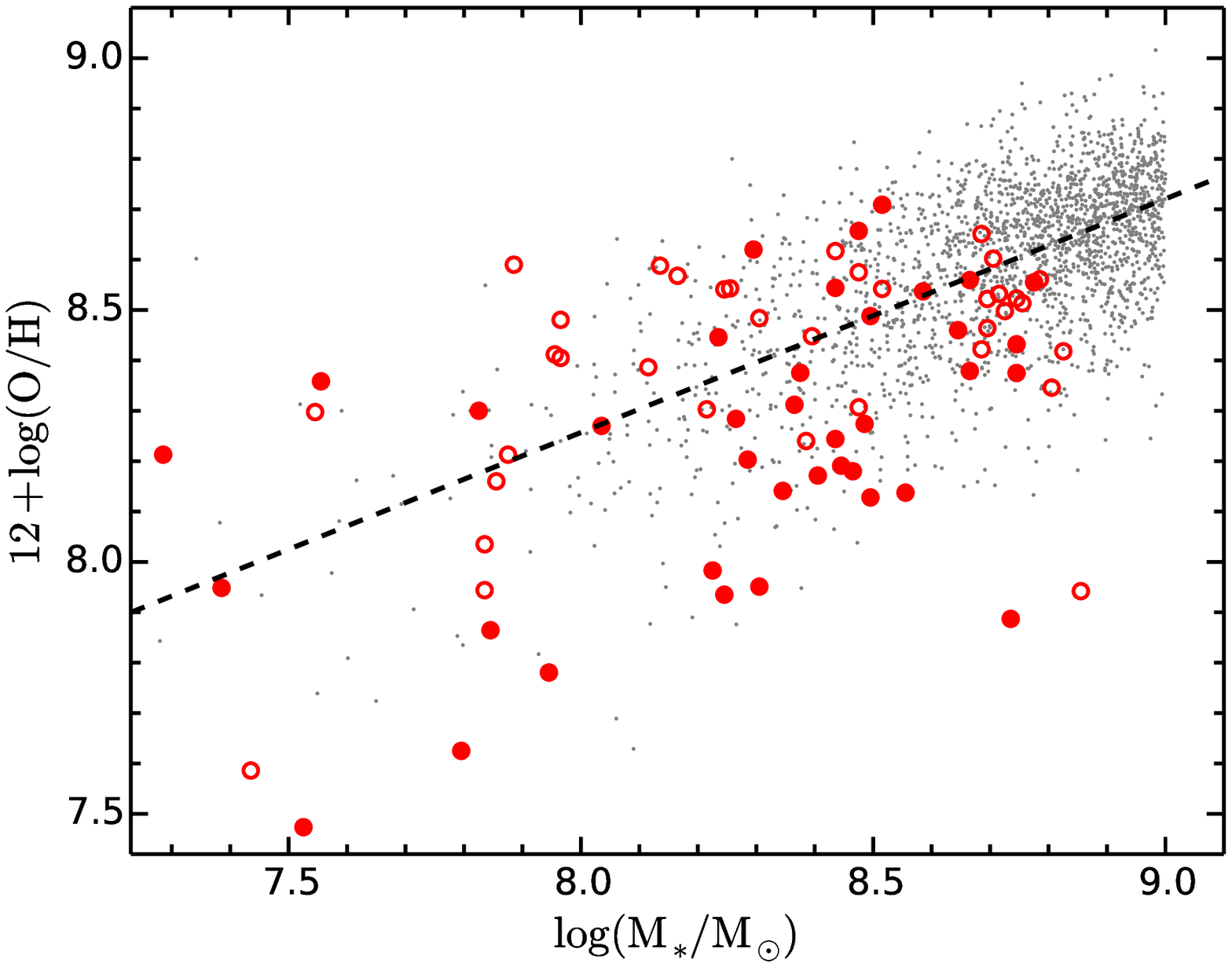}
 \caption{Mass-metallicity relation for BCDs. The symbols are the same as {Figure 3}.}
% while the COSMOS BCD sample is divided into 
% two redshift bins and separately shown in the two panels. }
\label{figure4}
\end{figure*}

{It is interesting that {the} COSMOS BCDs have metallicities 
{that are fairly} consistent with {those} of {the} local BCDs with {a} {slight} deviation of -0.05$\pm$0.16 dex.
Although {the} {intermediate-$z$} BCDs have R23 {ratios} {systematically higher} than {the} local BCDs,
the difference in metallicity is not significant. 
This is mainly because {intermediate-$z$} BCDs have intermediate metallicity where the R23 ratio 
is not a {monotonic} function of {the} metallicity. 
Meanwhile, the {intermediate-$z$} BCDs have O32 {ratios} {similar} to local BCDs which could
reduce the difference in {the} metallicities caused by different R23 {ratios}.
The consistent MZ relation between {the} {intermediate-$z$} and local BCDs {suggests} that the 
cosmic evolution of metallicity is not significant in {low-mass} galaxies from redshift z $\sim$ 0.3 to $\sim$ 0.03. 
This is not consistent with the result found by \citet{henry2013}. 
We consider {that} the different method {used} for metallicity determination is the main reason {for this inconsistency}. 
It should be noted that our sample {was} about three times larger and the method {of} metallicity determination 
adopted in the work {was} more consistent with other metallicity calibrations, {such as} the ${\rm T_{e}}$ method.
However, our metallicity measurements derived by the M08 method still {suffered} large uncertainties.
Information {on} \nii\ is important to derive metallicities with multiple calibrations 
to confirm or deny our result. These calibrations are monotonic with metallicity 
and do not have the {two-solution} problem.}

\subsection{Evolution of SFR and \dfn\ }

Besides metallicity, we also {obtained} the SFR and \dfn\ of BCDs.
Figure 5 shows BCDs in {the} mass-SFR and mass-\dfn\ diagram.
{The dashed lines represent the best-fitting relation to {the} SDSS BCDs.
It can be seen that intermediate-$z$ BCDs exhibit {a} consistent mass-SFR (main sequence) relation with local BCDs.
The median deviation of {the} COSMOS BCDs from the best-fitting line is 0.05 dex which is not significant compared {with}
the dispersion in {the} main sequence relation of 0.32 dex.
In contrast, the offset between the two BCD samples in {the} mass-\dfn\ relation seems to be larger,
being two times higher than the scatter.}%, is still not statistically significant.}}

{According to the downsizing evolution scenario, {low-mass} galaxies may experience star formation {that decreases less in the recent universe than in massive galaxies}. The insignificant deviation in SFR between {the} {intermediate-$z$} and local 
BCDs may be due to the large scatter in {the main sequence} relation and narrow redshift range.
However, \citet{Izotov2015} investigated aperture-corrected compact galaxy global parameters,
including stellar mass, gas-phase oxygen abundance, and SFR, in a wide range of redshift up to 3
and found weak redshift evolution. This is consistent with our results that no significant deviation in 
{MZ} and main sequence relation is found between intermediate-$z$ and local BCDs.
%of compact galaxy global parameters, including stellar mass, gas-phase oxygen abundance, SFR, in a wide range of 0-3. 
The lower \dfn\ of {the} intermediate-$z$ BCDs,
indicating {a} younger stellar population, is reasonable, since they {were} located in the {universe} 
$\sim$ 3 Gyrs earlier {than the local BCDs}.}

\begin{figure*}
 \centering
 \includegraphics[width=\textwidth]{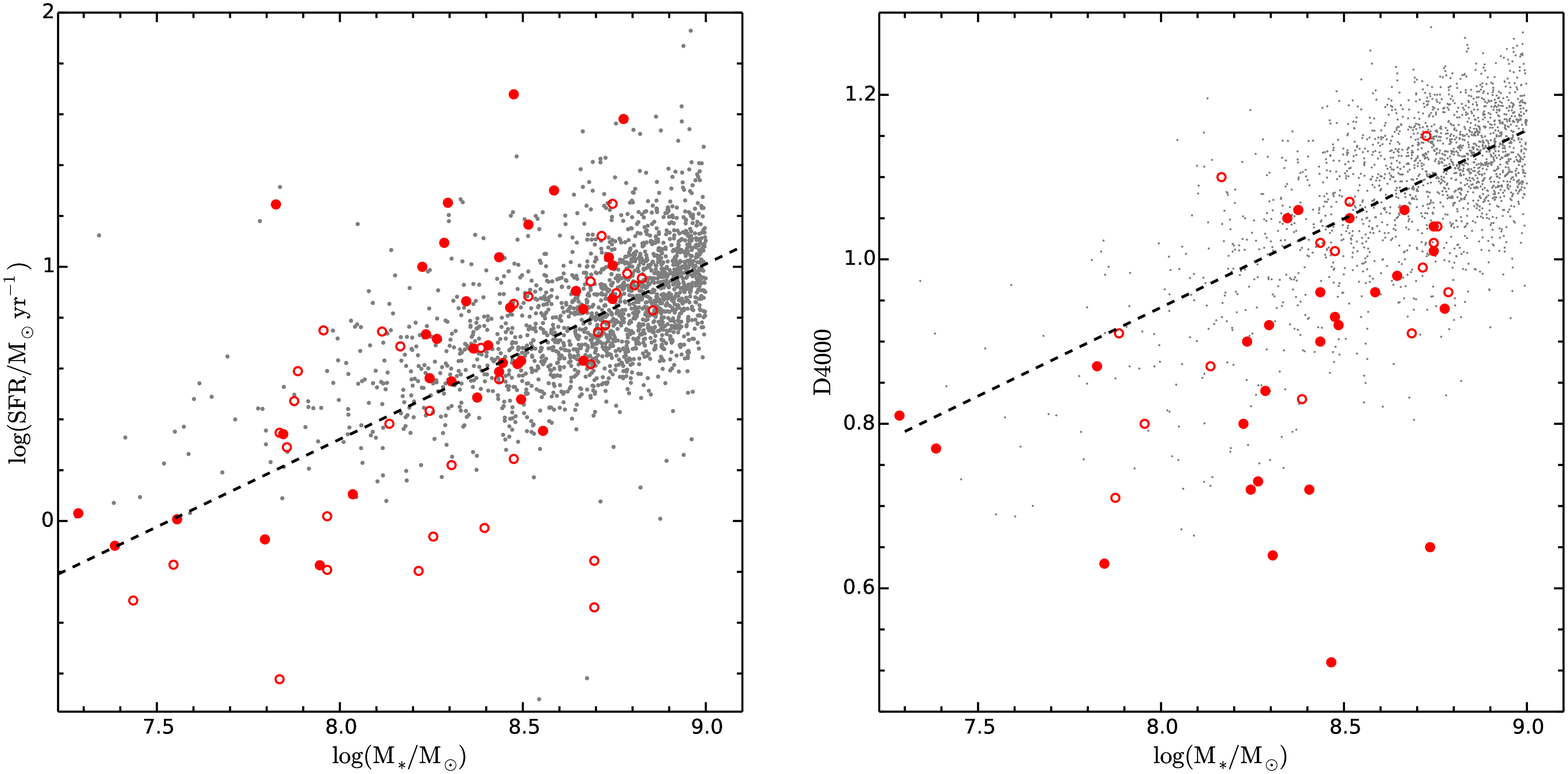}
 \caption{Stellar mass versus {SFR and \dfn\ } diagram for BCDs. Symbols are the same as Figure 3.}
\label{figure6}
\end{figure*}

\section{Summary}
In this work, we {have investigated} the {cosmic evolution of {the} metallicity, SFR, and \dfn\ } of {low-mass} {compact} {star-forming} galaxies from 
intermediate redshift to {the} local universe. 
We {selected} a local BCD sample from {the} SDSS database and {an} intermediate-$z$ BCD sample from 
{the} COSMOS {deep} field. Using Hectospec on {the} MMT, we obtained optical spectra for 180 COSMOS BCDs. 
Among them, {74} objects {had} reliable detection of strong emission lines, which {allows} metallicity 
determination {based on {the} strong-line method}. 
{Of the COSMOS BCD sample, more} than half
{had} reliable \hg\ detection and {had} dust {extinction} corrected {using the} 
Balmer decrement method. The main results and conclusions {were} as follows.

\begin{itemize}
 \item {{Intermediate-$z$ BCDs} have {R23 {ratios} systematically higher} than local BCDs, while {no 
 significant difference {was} found with the O32 ratio.}}
 %the O32 ratio of the two BCD samples seems to be similar.}

 \item {{We obtained oxygen abundance measurements for 74 COSMOS BCDs based on {the} strong-line method from \citet{maiolino2008}.
 Interestingly, the {MZ} relation of BCDs at {the} intermediate {redshift} is {fairly}} consistent with {that of the} local BCDs.
 with {a} slight deviation of -0.05$\pm$0.16 dex.}

 \item {In addition to metallicity, no significant deviation in {the} mass-SFR relation {was} found between the intermediate-$z$ and 
 local BCDs. On the other hand, the intermediate-$z$ BCDs {seemed} to be younger than the local BCDs, with lower \dfn\ index {values}. 
However, the deviation in \dfn\ index {values} is not statistically significant which may be due to the narrow {redshift} range.}
 %\item {The SFR of BCDs at z $\sim$ 0.34 tend to be higher than local BCDs by at least 0.4 dex. 
 %Meanwhile, the intermediate-redshift BCDs have younger stellar population age than local BCDs with lower \dfn\ index.}
\end{itemize}
 
%{If the insignificant cosmic evolution of metallicity is confirmed, the different behavior of SFR and metallicity 
%indicates that there could be some physical processes regulate the metal enrichment of a galaxy other than star formation activity.}
{The insignificant deviation in {MZ} and mass-SFR relation between intermediate-$z$ and local BCDs {indicates} that 
these relations between {the} global parameters of {low-mass} compact galaxies may be universal.}
{These results of {low-mass} {compact} galaxies {may} {provide}} important {observational}
constraints {onto} {the} {galaxy-formation} and evolution model.

\section*{Acknowledgments}
{We are grateful to referee's insightful suggestions and comments.}
We {thank} Yang, C. W. for his help in emission line measurements, Caldwell, N.
for his help in preparing observation and people in SAO Telescope Data Center for their help in data reduction. 
This research uses data obtained through the Telescope Access Program (TAP), which has been funded by the 
National Astronomical Observatories of China, the Chinese Academy of Sciences 
(the Strategic Priority Research Program `The Emergence of Cosmological Structures' Grant No. XDB09000000), 
and the Special Fund for Astronomy from the Ministry of Finance.
Hectospec observations reported here were obtained at the MMT Observatory, 
a joint facility of the University of Arizona and the Smithsonian Institution.
{This work is supported by the Strategic Priority Research Program `The Emergence of Cosmological Structures'
of the Chinese Academy of Sciences (No. XDB09000000), the Chinese National 973 Fundamental Science Programs (973 program) 
(2015CB857004), the National Natural Science Foundation of China (NSFC, Nos. 11225315, 1320101002,
11433005, 11421303, and 11303002), the Specialized Research Fund for the Doctoral Program of Higher Education (SRFDP,
No. 20123402110037),
the Yunnan Applied Basic Research Projects (2014FB155), and the Open Research
Program of Key Laboratory for Research in Galaxies and Cosmology, CAS.}

\clearpage
%\LongTables % optionally
% \begin{landscape}
  \begin{deluxetable}{lcccccccc}
  \tabletypesize{\small}
  \tablewidth{0pt}
  \rotate
  \tablecaption{{ Emission line measurements} of COSMOS BCD sample.}
%  \tablewidth{0pt}
%  \tabcolsep=0.16cm
%    \label{table:1}
%  \centering
%  \begin{tabular}{l c c c c c c c c c c c c}
  \tablehead{
  \colhead{ID} & \colhead{RA (J2000)} & \colhead{Dec (J2000)} & \colhead{\oii$\lambda$3727 $^a$}
  & \colhead{\hg} & \colhead{\hb} & \colhead{EW(\hb)} & \colhead{\oiii$\lambda$4959} & \colhead{\oiii$\lambda$5007} \\
  & \colhead{h:m:s} & \colhead{d:m:s} & & & & \colhead{\AA} & & 
  %& \colhead{$A_{\rm V}$ $^b$} & \colhead{log$M_*$} & \colhead{12+log(O/H)} & \colhead{log$q$} & \colhead{12+log(O/H)} \\
  %  & \colhead{h:m:s} & \colhead{d:m:s} & & & & & & & & mag & ${\rm M_{\odot}}$ 
  % & (R23) & cm/s & ($T_{\rm e}$) 
  }
  \rotate
  \startdata
1  &  09:59:37.39  &  +02:05:04.8  &  142.3$\pm$4.0  &  39.4$\pm$2.0  &  65.8$\pm$2.0  &  24.2  &  110.9$\pm$2.3  &  298.9$\pm$2.7  \\
2  &  09:59:15.82  &  +02:15:27.3  &  155.8$\pm$6.4  &  33.7$\pm$3.8  &  54.6$\pm$3.2  &  17.5  &  66.7$\pm$2.8  &  206.7$\pm$3.8  \\
3  &  09:59:45.49  &  +02:08:28.6  &  285.1$\pm$4.5  &  -  &  51.9$\pm$2.1  &  12.4  &  57.1$\pm$2.0  &  187.8$\pm$2.4  \\
4  &  09:59:27.85  &  +02:16:04.3  &  273.0$\pm$6.4  &  -  &  75.6$\pm$3.1  &  12.6  &  69.5$\pm$2.6  &  248.7$\pm$3.8  \\
5  &  09:59:19.76  &  +02:03:39.9  &  56.4$\pm$4.6  &  -  &  37.7$\pm$2.1  &  243.7  &  83.6$\pm$2.4  &  223.1$\pm$2.8  \\
6  &  09:58:32.84  &  +02:26:18.8  &  121.3$\pm$4.3  &  29.6$\pm$3.1  &  52.0$\pm$1.9  &  17.5  &  106.2$\pm$2.0  &  287.9$\pm$3.5  \\
7  &  09:59:31.40  &  +02:33:52.7  &  113.3$\pm$4.5  &  -  &  25.3$\pm$2.2  &  7.4  &  26.6$\pm$1.8  &  84.5$\pm$2.2  \\
8  &  09:59:09.46  &  +02:25:04.8  &  200.4$\pm$5.2  &  -  &  56.6$\pm$2.6  &  8.5  &  114.6$\pm$2.8  &  302.0$\pm$3.2  \\
9  &  09:59:15.11  &  +02:36:33.8  &  227.7$\pm$4.1  &  73.2$\pm$2.3  &  133.8$\pm$2.6  &  42.0  &  352.9$\pm$2.9  &  1010.0$\pm$4.1  \\
10  &  09:59:08.21  &  +02:26:00.9  &  94.7$\pm$5.0  &  -  &  29.1$\pm$2.9  &  13.0  &  25.9$\pm$2.2  &  68.3$\pm$2.7  \\
11  &  10:00:06.00  &  +02:34:53.7  &  135.6$\pm$5.4  &  -  &  26.2$\pm$2.7  &  8.2  &  21.9$\pm$2.5  &  76.4$\pm$2.6  \\
12  &  09:58:49.57  &  +02:04:56.9  &  188.9$\pm$6.6  &  77.8$\pm$4.5  &  138.6$\pm$3.7  &  104.5  &  229.4$\pm$4.1  &  689.9$\pm$4.8  \\
13  &  09:59:56.98  &  +02:12:54.9  &  91.7$\pm$5.0  &  -  &  44.4$\pm$3.9  &  48.3  &  40.0$\pm$2.8  &  101.6$\pm$2.8  \\
14  &  09:59:37.69  &  +02:09:02.9  &  35.2$\pm$4.2  &  -  &  56.7$\pm$4.0  &  85.8  &  98.7$\pm$3.4  &  270.0$\pm$4.4  \\
15  &  10:00:05.64  &  +02:16:08.0  &  4055.4$\pm$8.1  &  390.5$\pm$3.7  &  835.9$\pm$4.1  &  18.9  &  1002.0$\pm$4.0  &  2820.8$\pm$4.8  \\
16  &  09:59:37.27  &  +02:12:46.5  &  117.2$\pm$5.4  &  -  &  34.5$\pm$3.2  &  22.4  &  44.3$\pm$3.7  &  110.9$\pm$3.2  \\
17  &  09:58:44.54  &  +02:09:54.3  &  410.0$\pm$5.9  &  83.3$\pm$3.8  &  127.4$\pm$3.1  &  25.1  &  144.9$\pm$2.9  &  442.4$\pm$3.7  \\
18  &  09:58:59.75  &  +02:18:49.3  &  70.6$\pm$3.8  &  20.2$\pm$3.9  &  25.4$\pm$2.0  &  61.7  &  30.4$\pm$1.8  &  97.7$\pm$2.9  \\
19  &  09:59:26.78  &  +02:18:59.3  &  427.7$\pm$4.4  &  61.1$\pm$2.4  &  127.6$\pm$2.1  &  58.0  &  183.9$\pm$2.5  &  549.0$\pm$3.2  \\
20  &  09:58:51.38  &  +02:17:57.1  &  488.4$\pm$6.7  &  71.4$\pm$3.5  &  147.4$\pm$3.4  &  13.8  &  175.0$\pm$3.3  &  540.2$\pm$4.4  \\
21  &  09:59:54.00  &  +02:18:59.5  &  153.3$\pm$4.3  &  34.1$\pm$4.4  &  65.3$\pm$2.3  &  39.5  &  100.8$\pm$3.4  &  325.3$\pm$3.4  \\
22  &  09:59:26.41  &  +02:22:01.7  &  147.2$\pm$4.3  &  -  &  51.3$\pm$2.0  &  52.5  &  114.1$\pm$2.1  &  330.9$\pm$2.5  \\
23  &  09:58:31.33  &  +02:23:13.0  &  299.6$\pm$5.1  &  57.5$\pm$2.8  &  118.6$\pm$2.1  &  60.8  &  182.6$\pm$2.3  &  554.0$\pm$3.1  \\
24  &  09:59:45.54  &  +02:25:39.6  &  108.9$\pm$4.4  &  68.2$\pm$3.0  &  134.0$\pm$2.7  &  273.8  &  281.3$\pm$3.0  &  841.6$\pm$4.1  \\
25  &  09:59:06.68  &  +02:26:45.4  &  76.0$\pm$3.9  &  -  &  17.0$\pm$1.9  &  15.4  &  30.7$\pm$2.5  &  54.6$\pm$2.9  \\
26  &  09:59:30.22  &  +02:27:21.4  &  30.9$\pm$3.6  &  -  &  18.2$\pm$1.9  &  18.9  &  36.0$\pm$2.2  &  95.2$\pm$2.3  \\
27  &  09:58:53.41  &  +02:28:10.1  &  96.1$\pm$5.5  &  16.6$\pm$2.7  &  35.4$\pm$3.3  &  29.8  &  53.3$\pm$2.9  &  165.0$\pm$3.3  \\
28  &  09:59:07.98  &  +02:27:51.3  &  627.9$\pm$6.2  &  108.0$\pm$3.6  &  231.0$\pm$3.4  &  47.3  &  358.2$\pm$3.4  &  1051.0$\pm$4.7  \\
29  &  10:00:03.02  &  +02:22:21.4  &  165.5$\pm$3.9  &  -  &  35.3$\pm$1.8  &  45.6  &  37.8$\pm$2.0  &  108.7$\pm$1.9  \\
30  &  09:59:18.10  &  +02:34:33.3  &  242.3$\pm$4.1  &  -  &  52.3$\pm$2.1  &  19.2  &  44.8$\pm$1.9  &  148.7$\pm$2.2  \\
31  &  09:59:09.61  &  +02:35:47.6  &  157.2$\pm$3.5  &  -  &  37.8$\pm$1.6  &  203.3  &  8.1$\pm$1.4  &  130.6$\pm$2.1  \\
32  &  09:59:56.51  &  +02:26:25.5  &  117.4$\pm$3.5  &  28.9$\pm$2.5  &  48.7$\pm$1.9  &  73.3  &  97.0$\pm$2.2  &  282.9$\pm$2.7  \\
33  &  09:59:34.95  &  +02:27:56.2  &  55.6$\pm$4.3  &  -  &  29.7$\pm$3.2  &  43.9  &  23.9$\pm$2.3  &  71.8$\pm$3.0  \\
34  &  09:59:40.86  &  +02:28:47.1  &  104.8$\pm$4.7  &  -  &  24.1$\pm$2.5  &  27.1  &  25.3$\pm$2.6  &  77.2$\pm$3.2  \\
35  &  09:59:49.14  &  +02:37:00.1  &  213.2$\pm$5.3  &  64.2$\pm$3.5  &  123.0$\pm$2.5  &  66.6  &  253.1$\pm$2.7  &  798.7$\pm$4.0  \\
36  &  09:59:21.32  &  +02:38:54.4  &  287.9$\pm$6.2  &  -  &  69.5$\pm$3.1  &  10.2  &  80.1$\pm$3.2  &  238.6$\pm$3.5  \\
37  &  09:59:25.29  &  +02:39:16.3  &  132.8$\pm$5.5  &  -  &  44.4$\pm$2.9  &  50.0  &  54.1$\pm$3.3  &  142.1$\pm$5.4  \\
38  &  09:59:58.87  &  +02:29:47.6  &  330.2$\pm$4.5  &  55.2$\pm$3.0  &  106.0$\pm$2.2  &  63.8  &  156.3$\pm$2.6  &  423.4$\pm$3.1  \\
39  &  09:59:26.96  &  +02:46:05.4  &  276.6$\pm$3.6  &  47.8$\pm$2.2  &  102.4$\pm$2.1  &  50.8  &  201.9$\pm$2.2  &  592.3$\pm$2.7  \\
40  &  10:00:01.03  &  +02:47:35.7  &  203.3$\pm$6.6  &  68.2$\pm$3.7  &  113.6$\pm$3.9  &  76.4  &  153.3$\pm$3.6  &  428.3$\pm$4.1  \\
41  &  10:00:24.36  &  +02:27:14.9  &  195.2$\pm$4.0  &  37.1$\pm$3.2  &  66.6$\pm$2.8  &  26.7  &  70.5$\pm$2.0  &  208.3$\pm$2.6  \\
42  &  10:00:28.57  &  +02:33:53.2  &  151.8$\pm$5.3  &  31.9$\pm$3.1  &  42.5$\pm$3.0  &  16.0  &  55.0$\pm$2.8  &  141.4$\pm$2.9  \\
43  &  10:01:05.02  &  +02:41:04.9  &  1615.9$\pm$6.3  &  148.2$\pm$3.7  &  317.1$\pm$4.2  &  12.9  &  301.2$\pm$3.4  &  895.7$\pm$4.0  \\
44  &  10:01:23.48  &  +02:38:11.6  &  391.0$\pm$5.0  &  -  &  97.6$\pm$2.2  &  12.6  &  109.2$\pm$2.4  &  319.3$\pm$2.8  \\
45  &  10:00:51.99  &  +02:26:20.8  &  167.9$\pm$4.2  &  -  &  46.9$\pm$2.6  &  11.0  &  32.4$\pm$1.9  &  96.5$\pm$2.1  \\
46  &  10:01:46.06  &  +02:36:47.6  &  979.2$\pm$5.7  &  94.6$\pm$2.9  &  202.4$\pm$2.9  &  13.0  &  141.5$\pm$2.7  &  400.9$\pm$3.3  \\
47  &  10:01:35.13  &  +02:30:54.9  &  1893.1$\pm$7.6  &  194.2$\pm$4.2  &  415.6$\pm$4.2  &  15.1  &  489.0$\pm$3.7  &  1419.8$\pm$4.6  \\
48  &  10:01:32.79  &  +02:30:01.9  &  158.1$\pm$5.9  &  59.5$\pm$3.5  &  106.5$\pm$3.8  &  35.1  &  244.5$\pm$3.6  &  736.5$\pm$4.7  \\
49  &  10:02:08.14  &  +02:32:13.8  &  200.4$\pm$3.7  &  48.2$\pm$2.4  &  89.7$\pm$2.0  &  37.9  &  169.0$\pm$1.9  &  449.5$\pm$2.5  \\
50  &  10:02:05.01  &  +02:26:44.3  &  55.2$\pm$3.0  &  -  &  27.3$\pm$2.2  &  30.5  &  35.0$\pm$2.1  &  114.2$\pm$2.1  \\
51  &  10:01:57.48  &  +02:24:20.2  &  123.9$\pm$3.8  &  -  &  42.0$\pm$1.8  &  29.2  &  67.2$\pm$1.9  &  230.0$\pm$2.4  \\
52  &  10:01:30.91  &  +02:23:02.8  &  213.3$\pm$3.6  &  58.9$\pm$3.8  &  74.5$\pm$2.0  &  39.0  &  110.3$\pm$2.0  &  331.1$\pm$2.8  \\
53  &  10:01:10.73  &  +02:20:49.2  &  1455.2$\pm$9.4  &  444.9$\pm$5.1  &  924.1$\pm$5.3  &  88.6  &  1717.4$\pm$6.3  &  5022.1$\pm$9.7  \\
54  &  10:02:14.82  &  +02:19:36.0  &  94.9$\pm$3.7  &  -  &  40.8$\pm$1.9  &  22.7  &  44.7$\pm$2.1  &  126.0$\pm$2.0  \\
55  &  10:01:52.70  &  +02:14:15.9  &  75.0$\pm$6.2  &  95.7$\pm$4.2  &  160.7$\pm$3.7  &  91.1  &  438.7$\pm$4.0  &  1239.3$\pm$5.3  \\
56  &  10:00:53.37  &  +02:17:49.2  &  96.6$\pm$3.6  &  24.3$\pm$2.3  &  42.6$\pm$1.9  &  43.7  &  78.6$\pm$2.4  &  237.7$\pm$2.6  \\
57  &  10:01:34.72  &  +02:13:30.3  &  271.5$\pm$4.3  &  -  &  58.4$\pm$2.0  &  10.8  &  53.2$\pm$2.2  &  209.0$\pm$3.9  \\
58  &  10:01:41.88  &  +02:06:25.1  &  159.8$\pm$5.1  &  -  &  37.6$\pm$2.0  &  17.5  &  51.8$\pm$2.1  &  173.4$\pm$4.2  \\
59  &  10:00:57.40  &  +02:09:05.8  &  73.6$\pm$3.9  &  -  &  29.7$\pm$2.0  &  101.2  &  40.9$\pm$1.9  &  119.4$\pm$2.2  \\
60  &  10:01:15.57  &  +02:04:00.3  &  309.8$\pm$5.7  &  40.4$\pm$4.2  &  83.0$\pm$3.2  &  19.5  &  76.2$\pm$2.8  &  231.5$\pm$3.1  \\
61  &  10:00:42.57  &  +02:12:20.4  &  90.7$\pm$5.7  &  37.0$\pm$3.2  &  78.9$\pm$3.4  &  109.4  &  169.7$\pm$3.5  &  508.2$\pm$4.2  \\
62  &  10:01:20.81  &  +01:54:51.5  &  156.6$\pm$3.8  &  -  &  54.0$\pm$1.9  &  32.2  &  58.2$\pm$2.0  &  183.7$\pm$2.8  \\
63  &  10:00:46.26  &  +01:55:48.5  &  296.2$\pm$4.0  &  -  &  93.7$\pm$2.2  &  41.2  &  133.7$\pm$2.1  &  414.2$\pm$2.7  \\
64  &  10:00:18.29  &  +02:03:11.8  &  121.8$\pm$4.3  &  -  &  26.3$\pm$1.8  &  147.8  &  43.6$\pm$2.7  &  113.6$\pm$2.3  \\
65  &  10:00:28.31  &  +02:01:24.7  &  3586.7$\pm$5.4  &  295.3$\pm$3.4  &  632.1$\pm$2.3  &  30.4  &  653.2$\pm$2.4  &  1659.3$\pm$2.9  \\
66  &  09:59:41.99  &  +01:58:17.7  &  183.5$\pm$4.8  &  36.0$\pm$3.2  &  77.0$\pm$2.2  &  36.8  &  116.7$\pm$2.3  &  320.7$\pm$2.6  \\
67  &  10:00:06.45  &  +02:09:14.0  &  220.8$\pm$5.4  &  78.6$\pm$3.3  &  156.3$\pm$2.6  &  71.8  &  290.3$\pm$3.0  &  848.5$\pm$4.0  \\
68  &  09:59:28.49  &  +01:56:47.7  &  333.5$\pm$6.0  &  -  &  84.7$\pm$3.2  &  15.2  &  69.7$\pm$3.0  &  240.6$\pm$3.1  \\
69  &  10:02:08.43  &  +02:27:21.4  &  59.3$\pm$2.9  &  -  &  60.0$\pm$1.9  &  35.6  &  84.9$\pm$2.0  &  216.5$\pm$2.6  \\
70  &  10:01:12.67  &  +02:34:14.8  &  854.3$\pm$5.5  &  111.0$\pm$3.2  &  237.5$\pm$2.9  &  23.2  &  200.8$\pm$2.6  &  678.9$\pm$3.8  \\
71  &  10:01:46.92  &  +02:41:50.9  &  496.1$\pm$5.1  &  87.8$\pm$2.8  &  167.9$\pm$2.4  &  28.0  &  272.4$\pm$3.2  &  785.2$\pm$3.9  \\
72  &  10:01:50.75  &  +02:26:49.8  &  243.9$\pm$4.1  &  -  &  41.3$\pm$1.8  &  26.6  &  46.0$\pm$2.0  &  143.4$\pm$2.3  \\
73  &  09:59:57.35  &  +01:52:17.9  &  57.7$\pm$3.6  &  -  &  24.0$\pm$1.8  &  14.7  &  26.6$\pm$1.9  &  72.0$\pm$2.0  \\
74  &  10:00:35.77  &  +02:16:33.3  &  72.9$\pm$5.6  &  -  &  27.7$\pm$3.9  &  30.0  &  31.8$\pm$2.3  &  117.7$\pm$3.3  \\
\enddata
\tablenotetext{a}{All the emission lines have units of $10^{-18} {\rm erg\ s^{-1}\ cm^{-2}}$ and {
EW} of \AA.}
%\tablenotetext{b}{The dust attenuation $A_{\rm V}$ from SED fitting is listed within parantheses.}
 \end{deluxetable}
%\end{landscape}

\begin{deluxetable}{lcccccc}
  \tabletypesize{\small}
  \tablewidth{0pt}
  \rotate
  \tablecaption{{Physical properties of COSMOS BCD sample.}}
%  \tablewidth{0pt}
%  \tabcolsep=0.16cm
%    \label{table:1}
%  \centering
%  \begin{tabular}{l c c c c c c c c c c c c}
  \tablehead{
  \colhead{ID} %& \colhead{RA (J2000)} & \colhead{Dec (J2000)} & \colhead{\oii$\lambda$3727 $^a$}
%  & \colhead{\hg} & \colhead{\hb} & \colhead{EW(\hb)} & \colhead{\oiii$\lambda$4959} & \colhead{\oiii$\lambda$5007} 
  & \colhead{$z$} & \colhead{$A_{\rm V}$ $^a$} & \colhead{log$M_*$} & \colhead{12+log(O/H)} & \colhead{log(SFR) $^b$} & \colhead{\dfn} \\
  & & mag & ${\rm M_{\odot}}$ & & ${\rm M_{\odot}\ yr^{-1}}$ & 
%    & \colhead{h:m:s} & \colhead{d:m:s} & & & & & & & & mag & ${\rm M_{\odot}}$
  }
  \rotate
  \startdata
1  &  0.482  &  0.0  &  8.29  &  8.20$\pm$0.04  &  0.88$\pm$0.03  &  0.8  \\
2  &  0.250  &  0.0  &  8.38  &  8.38$\pm$0.06  &  0.16$\pm$0.01  &  1.1  \\
3  &  0.358  &  (0.4)  &  8.79  &  8.56$\pm$0.03  &  0.34$\pm$0.01  &  1.0  \\
4  &  0.250  &  (0.1)  &  8.73  &  8.50$\pm$0.03  &  0.22$\pm$0.01  &  1.1  \\
5  &  0.305  &  (0.0)  &  7.84  &  7.94$\pm$0.13  &  0.17$\pm$0.01  &  -  \\
6  &  0.372  &  0.0  &  8.35  &  8.14$\pm$0.06  &  0.37$\pm$0.01  &  1.1  \\
7  &  0.356  &  (0.0)  &  8.52  &  8.54$\pm$0.06  &  0.17$\pm$0.01  &  1.1  \\
8  &  0.261  &  (0.2)  &  8.48  &  8.31$\pm$0.04  &  0.18$\pm$0.01  &  1.0  \\
9  &  0.427  &  0.0  &  8.74  &  7.89$\pm$0.03  &  1.34$\pm$0.03  &  0.7  \\
10  &  0.261  &  (0.0)  &  8.14  &  8.59$\pm$0.08  &  0.09$\pm$0.01  &  0.9  \\
11  &  0.266  &  (0.1)  &  8.44  &  8.62$\pm$0.06  &  0.09$\pm$0.01  &  1.0  \\
12  &  0.202  &  0.0  &  7.39  &  7.95$\pm$0.06  &  0.24$\pm$0.01  &  0.8  \\
13  &  0.267  &  (0.0)  &  7.97  &  8.48$\pm$0.09  &  0.15$\pm$0.01  &  -  \\
14  &  0.221  &  (0.0)  &  7.44  &  7.59$\pm$0.18  &  0.12$\pm$0.01  &  -  \\
15  &  0.236  &  1.4  &  8.78  &  8.55$\pm$0.00  &  2.11$\pm$0.01  &  0.9  \\
16  &  0.261  &  (0.0)  &  8.31  &  8.48$\pm$0.07  &  0.11$\pm$0.01  &  0.7  \\
17  &  0.260  &  0.0  &  8.24  &  8.45$\pm$0.02  &  0.40$\pm$0.01  &  0.9  \\
18  &  0.372  &  0.0  &  7.56  &  8.36$\pm$0.08  &  0.18$\pm$0.01  &  -  \\
19  &  0.426  &  0.0  &  8.75  &  8.38$\pm$0.01  &  1.27$\pm$0.02  &  1.0  \\
20  &  0.246  &  0.0  &  8.75  &  8.43$\pm$0.02  &  0.41$\pm$0.01  &  1.0  \\
21  &  0.377  &  0.0  &  8.45  &  8.19$\pm$0.04  &  0.49$\pm$0.02  &  -  \\
22  &  0.340  &  (0.4)  &  7.86  &  8.16$\pm$0.04  &  0.30$\pm$0.01  &  -  \\
23  &  0.339  &  0.0  &  8.44  &  8.24$\pm$0.02  &  0.69$\pm$0.01  &  0.9  \\
24  &  0.320  &  0.0  &  7.80  &  7.62$\pm$0.06  &  0.68$\pm$0.01  &  -  \\
25  &  0.372  &  (0.1)  &  8.25  &  8.54$\pm$0.08  &  0.12$\pm$0.01  &  -  \\
26  &  0.362  &  (0.0)  &  7.84  &  8.04$\pm$0.19  &  0.12$\pm$0.01  &  -  \\
27  &  0.260  &  0.3  &  8.04  &  8.27$\pm$0.08  &  0.11$\pm$0.01  &  0.8  \\
28  &  0.260  &  0.4  &  8.27  &  8.28$\pm$0.01  &  0.72$\pm$0.01  &  0.7  \\
29  &  0.359  &  (0.3)  &  8.48  &  8.57$\pm$0.04  &  0.23$\pm$0.01  &  -  \\
30  &  0.343  &  (0.1)  &  8.71  &  8.60$\pm$0.03  &  0.31$\pm$0.01  &  1.0  \\
31  &  0.444  &  (0.1)  &  8.70  &  8.52$\pm$0.03  &  0.42$\pm$0.02  &  -  \\
32  &  0.426  &  0.0  &  8.56  &  8.14$\pm$0.04  &  0.49$\pm$0.02  &  -  \\
33  &  0.247  &  (0.0)  &  7.97  &  8.40$\pm$0.13  &  0.08$\pm$0.01  &  -  \\
34  &  0.250  &  (0.0)  &  8.26  &  8.54$\pm$0.07  &  0.07$\pm$0.01  &  -  \\
35  &  0.335  &  0.0  &  8.31  &  7.95$\pm$0.04  &  0.69$\pm$0.01  &  0.6  \\
36  &  0.269  &  (0.1)  &  8.76  &  8.51$\pm$0.03  &  0.24$\pm$0.01  &  1.0  \\
37  &  0.259  &  (0.1)  &  8.40  &  8.45$\pm$0.07  &  0.14$\pm$0.01  &  -  \\
38  &  0.379  &  0.0  &  8.67  &  8.38$\pm$0.02  &  0.80$\pm$0.02  &  -  \\
39  &  0.428  &  0.5  &  8.47  &  8.18$\pm$0.02  &  1.03$\pm$0.02  &  0.5  \\
40  &  0.219  &  0.0  &  7.29  &  8.21$\pm$0.05  &  0.24$\pm$0.01  &  0.8  \\
41  &  0.417  &  0.0  &  8.65  &  8.46$\pm$0.03  &  0.63$\pm$0.03  &  1.0  \\
42  &  0.266  &  0.0  &  8.50  &  8.49$\pm$0.05  &  0.14$\pm$0.01  &  -  \\
43  &  0.220  &  1.4  &  8.30  &  8.62$\pm$0.01  &  0.68$\pm$0.01  &  0.9  \\
44  &  0.360  &  (0.1)  &  8.75  &  8.52$\pm$0.02  &  0.65$\pm$0.01  &  1.0  \\
45  &  0.344  &  (0.2)  &  8.69  &  8.65$\pm$0.04  &  0.28$\pm$0.02  &  0.9  \\
46  &  0.246  &  0.8  &  8.52  &  8.71$\pm$0.01  &  0.56$\pm$0.01  &  1.1  \\
47  &  0.219  &  0.9  &  8.59  &  8.54$\pm$0.01  &  0.89$\pm$0.01  &  1.0  \\
48  &  0.219  &  0.0  &  7.85  &  7.86$\pm$0.06  &  0.23$\pm$0.01  &  0.6  \\
49  &  0.346  &  0.0  &  8.41  &  8.17$\pm$0.03  &  0.55$\pm$0.01  &  0.7  \\
50  &  0.422  &  (0.0)  &  7.88  &  8.21$\pm$0.09  &  0.26$\pm$0.02  &  0.7  \\
51  &  0.422  &  (0.1)  &  8.39  &  8.24$\pm$0.04  &  0.41$\pm$0.02  &  0.8  \\
52  &  0.375  &  0.0  &  8.37  &  8.31$\pm$0.02  &  0.55$\pm$0.01  &  -  \\
53  &  0.248  &  0.0  &  8.23  &  7.98$\pm$0.01  &  2.60$\pm$0.01  &  0.8  \\
54  &  0.410  &  (0.4)  &  8.12  &  8.39$\pm$0.06  &  0.37$\pm$0.02  &  -  \\
55  &  0.216  &  0.0  &  7.53  &  7.47$\pm$0.06  &  0.33$\pm$0.01  &  0.6  \\
56  &  0.474  &  0.0  &  8.50  &  8.13$\pm$0.06  &  0.55$\pm$0.02  &  -  \\
57  &  0.371  &  (0.3)  &  8.72  &  8.53$\pm$0.03  &  0.42$\pm$0.01  &  1.0  \\
58  &  0.465  &  (0.3)  &  8.83  &  8.42$\pm$0.05  &  0.46$\pm$0.02  &  -  \\
59  &  0.352  &  (0.0)  &  8.22  &  8.30$\pm$0.08  &  0.19$\pm$0.01  &  -  \\
60  &  0.310  &  0.0  &  8.67  &  8.56$\pm$0.03  &  0.39$\pm$0.02  &  1.1  \\
61  &  0.244  &  0.0  &  7.95  &  7.78$\pm$0.10  &  0.21$\pm$0.01  &  -  \\
62  &  0.373  &  (0.0)  &  8.69  &  8.42$\pm$0.04  &  0.39$\pm$0.01  &  -  \\
63  &  0.442  &  (0.3)  &  8.81  &  8.35$\pm$0.02  &  1.02$\pm$0.02  &  -  \\
64  &  0.379  &  (0.2)  &  8.70  &  8.46$\pm$0.05  &  0.20$\pm$0.01  &  -  \\
65  &  0.361  &  1.7  &  8.48  &  8.66$\pm$0.00  &  4.25$\pm$0.02  &  0.9  \\
66  &  0.339  &  0.4  &  8.49  &  8.27$\pm$0.04  &  0.45$\pm$0.01  &  0.9  \\
67  &  0.316  &  0.0  &  8.25  &  7.93$\pm$0.04  &  0.77$\pm$0.01  &  0.7  \\
68  &  0.238  &  (0.2)  &  8.17  &  8.57$\pm$0.03  &  0.22$\pm$0.01  &  1.1  \\
69  &  0.456  &  (0.1)  &  8.86  &  7.94$\pm$0.11  &  0.70$\pm$0.02  &  0.5  \\
70  &  0.259  &  1.0  &  8.44  &  8.54$\pm$0.01  &  0.74$\pm$0.01  &  1.0  \\
71  &  0.401  &  0.0  &  7.83  &  8.30$\pm$0.01  &  1.45$\pm$0.02  &  0.9  \\
72  &  0.375  &  (0.4)  &  7.89  &  8.59$\pm$0.03  &  0.30$\pm$0.01  &  0.9  \\
73  &  0.428  &  (0.0)  &  7.96  &  8.41$\pm$0.09  &  0.24$\pm$0.02  &  0.8  \\
74  &  0.220  &  (0.0)  &  7.55  &  8.30$\pm$0.12  &  0.06$\pm$0.01  &  -  \\
\enddata
%\tablenotetext{a}{All the emission lines have units of $10^{-18} {\rm erg\ s^{-1}\ cm^{-2}}$ and equivalent width of \AA.}
\tablenotetext{a}{The dust attenuation $A_{\rm V}$ from SED fitting is listed within parantheses.}
\tablenotetext{b}{The listed SFR are not corrected for burst age which can be done by comining the EW(\hb) listed in Table 1 and 
{equation (5)} in \citet{Izotov2015}.}
 \end{deluxetable}
  
\end{document}